\documentclass[12pt]{article}
\usepackage{amsfonts,amssymb}

\def\bq{ \begin{equation} }
\def\eq{ \end{equation} }
\def\ben{ \begin{eqnarray} }
\def\en{ \end{eqnarray} }

\def\frac#1#2{{#1\over #2}}

\def\on#1#2{\mathop{\vbox{\ialign{##\crcr\noalign{\kern2pt}
$\scriptstyle{#2}$\crcr\noalign{\kern2pt\nointerlineskip}
\kern-2pt$\hfil\displaystyle{#1}\hfil$\crcr}}}\limits}

\begin{document}

\baselineskip=15pt
\vspace{1cm} \centerline{{\LARGE \textbf {Integrable
equations
 }}}
\vspace{0.3cm} \centerline{{\LARGE \textbf {and classical S-matrix
 }}}

\vskip1cm \hfill
\begin{minipage}{13.5cm}
\baselineskip=15pt {\bf V.E Zakharov ${}^{1,}{}^{2, }{}^{3}$,  A.V Odesskii
${}^{4}$, M. Onorato ${}^{5}$, M. Cisternino ${}^{5}$}
\\ [2ex] {\footnotesize
${}^{1}$  University Of Arizona, Tucson, USA
\\
${}^{2}$  Lebedev Physical Institute, Moscow, Russia
\\
${}^{3}$ Novosibirsk State University, Russia
\\
${}^4$ Brock University, St. Catharines, Canada
\\
${}^{5}$ University of Torino, Italy}
\vskip1cm{\bf Abstract}

We study amplitudes of five-wave interactions for evolution Hamiltonian equations differ from the KdV equation by 
the form of dispersion law. We find that five-wave amplitude is canceled for all three known equations (KdV, Benjamin-Ono and equation of intermediate waves) and 
for two new equations which are natural generalizations of mentioned above.

\end{minipage}

\vskip0.8cm \noindent{ }
\vglue1cm 

\textbf{E-mail}: zakharov@math.arizona.edu, aodesski@brocku.ca

\newpage \tableofcontents
\newpage

\section{Introduction}
\bigskip

At least three important Hamiltonian evolutionary equations that appear in the theory of ocean waves are completely integrable (see, for instance \cite{abl}). They are:

1. The KdV equation
\begin{equation}\label{kdv}
u_t=u_{xxx}+uu_x
\end{equation}

2. The Benjamin-Ono equation
\begin{equation}\label{bo}
u_t=\hat{I}(u_{xx})+uu_x
\end{equation}
Here $\hat{I}$ is the Hilbert transform.

3. The intermediate wave equation
\begin{equation}\label{eqgen}
u_t=\hat{F}(u)+uu_x
\end{equation}
Here $\hat{F}$ is a pseudo-differential operator with symbol
\begin{equation}\label{intw}
F(k)=ak^2\coth bk-ck. 
\end{equation}
In the limit $b\to\infty,~a=1,~c=0$ equation (\ref{intw}) tends to the Benjamin-Ono equation. In the limit $b\to 0,~a=\frac{3}{b},~c=\frac{3}{b^2}$ equation (\ref{intw}) goes to the KdV equation.

In this article we address the following question\footnote{Similar problem was discussed in \cite{mikhnov, honenov} in the framework of the symmetry approach}: can one find other integrable equations of the type (\ref{eqgen})? We assume that the discussed equations are Hamiltonian and admit the 
Gardner Poisson structure
$$u_t=\frac{\partial}{\partial x} \frac{\delta H}{\delta u}$$
or, in terms of Fourier transforms
$$u(k)_t=ik\frac{\delta H}{\delta u(k)},~~~u(k)=\frac{1}{2\pi}\int_{-\infty}^{\infty}u(x,t)e^{-ikx}dx$$

Here $H=H_2+H_3$ where
$$H_2=\frac{1}{2}\int_{-\infty}^{\infty}\frac{F(k)}{k}u(k)u(-k)dk,~~~H_3=\frac{1}{6}u^3=\frac{1}{6}\int_{-\infty}^{\infty}u(k_1)u(k_2)u(k_3)\delta(k_1+k_2+k_3)dk_1dk_2dk_3$$

Thus we assume that $F(k)$ is an odd function, $F(-k)=-F(k)$. For KdV we have $F(k)=-k^3$ and for  the Benjamin-Ono equation $F(k)=-|k|k$.

In this article we classify all integrable equations of the form (\ref{eqgen}). The answer is the following: there is only one extra equation given by 
\begin{equation}\label{cot}
F(k)=ak^2\cot bk-ck.
\end{equation}
In the limit $b\to 0,~a=-\frac{3}{b},~c=\frac{3}{b^2}$ we get $F(k)=-k^3$. The dispersion relation (\ref{cot}) has singularities at $k_n=\frac{\pi n}{b}$.

We also show that the following $(1+2)$-dimensional equation 
\begin{equation}\label{1+2dim}
\frac{\partial u}{\partial t} = \frac{\partial^2}{\partial x^2} \hat{L}\Big(\frac{\partial}{\partial y}\Big)u+uu_x
\end{equation}
is integrable, where $u=u(x,y,t)$ and $\hat{L}(p)=\epsilon\frac{e^{\epsilon p}+1}{e^{\epsilon p}+1}$. In the limit $\epsilon\to 0$ this equation reads: 
$$\frac{\partial u}{\partial t} = \frac{\partial^2}{\partial x^2} \Big(\frac{\partial}{\partial y}\Big)^{-1}u+uu_x.$$
Note that this equation is similar to  well-known Khokhlov-Zabolotskaya 
equation (see for instance \cite{z6}). Recall that the Khokhlov-Zabolotskaya  equation is the dispersionless limit of both KP1 and KP2 equations.

Note that equation \ref{eqgen} has the following universal conservation laws for an arbitrary function $F(k)$:
$$I_0=\int_{-\infty}^{\infty}udx,~I_1=\int_{-\infty}^{\infty}u^2dx,~I_3=H.$$
We pose the following question: for which functions $F(k)$ does there exist at least one additional conservation law given by a power series in $u$ starting with a quadratic term 
$$I_3=I^{(2)}+I^{(3)}+...$$
where
$$I^{(2)}=\int_{-\infty}^{\infty}g^{(2)}(k)u(k)u(-k)dk,~~~I^{(3)}=\int_{-\infty}^{\infty}g^{(3)}(k_1,k_2)u(k_1)u(k_2)u(-k_1-k_2)dk_1dk_2,...$$
and $g^{(2)}(k)=g^{(2)}(-k)$ is a real function that is not a linear combination $c_1k+c_2F(k)$?

The existence of such a conservation law is not a proof of integrability, while nonexistence is a clear manifestation of non-integrability. Thus to accomplish our task we must prove integrability of 
all new equations separately. We plan to do this in a future publication.

\section{Scattering matrix}
\bigskip

Following Zakharov and Shulman \cite{z1}-\cite{z5} we introduce a so-called formal scattering matrix for the equation (\ref{eqgen}) with arbitrary $F(k)$. We write this equation in 
Fourier components
\begin{equation}\label{four}
u(k)_t=iF(k)u(k)+ik\int_{-\infty}^{\infty}u(k_1)u(k_2)\delta(k-k_1-k_2)dk_1dk_2
\end{equation}
We introduce $c(k)$ by $u(k)=c(k)e^{iF(k)t}$, assuming that $c(k)\to c^{-}(k)$ when $t\to -\infty$, and rewrite equation (\ref{four}) in Picard form as follows
\begin{equation}\label{four1}
c(k)=c^{-}(k)+ik\lim_{\epsilon\to 0}\int_{-\infty}^{t}\int_{-\infty}^{\infty}c(k_1,\tau)c(k_2,\tau)e^{i(F(k_1)+F(k_2)-F(k))\tau-\epsilon|\tau|}\delta(k-k_1-k_2)dk_1dk_2d\tau
\end{equation}

We now solve equation (\ref{four1}) by iterations, sending $t\to\infty$ and then $\epsilon\to 0$. Let $c(k,t)\to c^{+}(k)$ when $t\to\infty$. We end up with $c^{+}$ expressed through $c^{-}$ in 
terms of the so-called formal scattering matrix $S$:
\begin{equation}\label{S}
c^{+}(k)=Sc^{-}(k)=c^{-}(k)+
\end{equation}
$$\sum_{n=2}^{\infty}\int_{-\infty}^{\infty}S(k,k_1,...,k_{n})\delta(F(k)-F(k_1)-...-F(k_{n}))\delta(k-k_1-...-k_{n})c^{-}(k_1)...c^{-}(k_{n})dk_1...dk_{n}$$
The functions $S(k,k_1,...,k_{n})$ are called amplitudes of wave scattering of order $n+1$. The arguments of delta-functions in (\ref{S}) are called resonance conditions and equation
$$S(k,k_1,...,k_{n})=0,$$ where $k,k_1,...,k_{n}$ are subject to resonance conditions, is called the $n+1$-wave equation. The first resonance condition 
$$F(k)=F(k_1)+F(k_2),~~~k=k_1+k_2$$
only has trivial solutions such as $k=0,~k_2=-k_1$ or $k_2=0,~k=k_1$. Therefore the three-wave equation is not significant. In the same way the four-wave resonance conditions 
$$F(k)=F(k_1)+F(k_2)+F(k_3),~~~k=k_1+k_2+k_3$$
have only trivial solutions such as $k=k_1,~k_3=-k_2$. Hence the first nonlinear resonance process is five-wave interaction, governed by resonance conditions
\begin{equation}\label{5w}
k=k_1+k_2+k_3+k_4,~~~F(k)=F(k_1)+F(k_2)+F(k_3)+F(k_4)
\end{equation}
Suppose that $k>0$. At least one wave vector in the right hand side of (\ref{5w}) must be negative. Assume that $k_4<0$. We replace $k_4\to-k_4$ and $k\to k_5$ and rewrite equations (\ref{5w}) as follows
\begin{equation}\label{5w1}
k_4+k_5=k_1+k_2+k_3,~~~F(k_4)+F(k_5)=F(k_1)+F(k_2)+F(k_3)
\end{equation}
All wave vectors in (\ref{5w1}) are positive. Moreover, we assume them ordered as follows
$$k_2>k_4>k_5>k_3>k_1$$
Under this assumption the five-wave amplitude is
$$S(k_1,k_2,k_3,k_4,k_5)=F_{12}(F_{45}+G_{53}+G_{43})+F_{13}(F_{45}+G_{23}+G_{24})+$$
$$G_{51}(F_{23}+G_{43}+G_{24})+G_{41}(F_{23}+G_{53}+G_{25})+F_{45}F_{23}+G_{24}G_{53}+G_{25}G_{43}$$
Here
$$F_{ij}=\frac{k_i+k_j}{F(k_i+k_j)-F(k_i)-F(k_j)},~~~G_{ij}=\frac{k_i-k_j}{F(k_i-k_j)-F(k_i)+F(k_j)}$$
for $i\ne j=1,...,5$.

The necessary condition for integrability is the cancellation of the five-wave amplitude on the resonance manifold (\ref{5w1}).

\section{Cancellation of five-wave amplitude for known integrable systems}
\bigskip

Let $F(k)=k^3$ (this is the KdV case). Then
$$F_{ij}=\frac{1}{3k_ik_j},~G_{ij}=-\frac{1}{3k_ik_j}.$$ After a simple calculation we obtain
\begin{equation}\label{skdv}
S_{12345}=\frac{1}{9k_1k_2k_3k_4k_5}(k_4+k_5-k_1-k_2-k_3)=0.
\end{equation}

Notice that to check the cancellation of the five-waves amplitude in this case we do not use the frequency resonance condition in (\ref{5w1}).

As long as all $k_i>0$ we can set $F(k)=k^2$ for the Benjamin-Ono case. Then
$$F_{ij}=\frac{k_i+k_j}{2k_ik_j},~~~G_{ij}=-\frac{1}{2k_j},$$
Now $$
S_{12345}=\frac{k_1k_4k_3+k_1k_5k_3-k_1k_4k_5+k_2k_4k_3+k_2k_5k_3-k_2k_4k_5-k_5k_3k_4+k_2k_1k_4+k_2k_1k_5}{2k_1k_2k_3k_4k_5}.
$$
This expression can be written in the form
\begin{equation}\label{sbo}
\frac{k_4+k_5}{4k_1k_2k_3k_4k_5}(k_4^2+k_5^2-k_1^2-k_2^2-k_3^2)+\frac{(k_1+k_2+k_3)(k_4+k_5)+k_4^2+k_5^2}{4k_1k_2k_3k_4k_5}(k_1+k_2+k_3-k_4-k_5).
\end{equation}
Thus the cancellation by virtue of (\ref{5w1}) is obvious.

To check cancellation for the generic dispersion relation (\ref{intw}), we first notice that $S_{12345}$ is invariant with respect to the transformation $F(k)\to F(\alpha k)+\beta k$, where 
$\alpha\ne 0$, $\beta$ are 
arbitrary constants. Moreover, we can replace $F(k)$ by $F(k,p)=k^2\frac{1+e^p}{1-e^p}$ because the exponential is not an algebraic function. The resonance conditions now read:
$$p_4+p_5=p_1+p_2+p_3,~~~k_4+k_5=k_1+k_2+k_3$$
and
$$F(k_4,p_4)+F(k_5,p_5)=F(k_1,p_1)+F(k_2,p_2)+F(k_3,p_3).$$
The five-wave amplitude depends on ten variables
$$S_{12345}=S(k_1,...,k_5,p_1,...,p_5).$$
Checking the cancellation of this amplitude by virtue of the resonance conditions took approximately ten minutes for Maple. The explicit representation of $S_{12345}$ in the form similar to (\ref{skdv}) and (\ref{sbo}) is so cumbersome that we do not present it here.

\section{Solving the functional equation}
\bigskip

Consider the five-wave equation $S(k_1,...,k_5)=0$ as a functional equation for the function $f(k)=F(k)$.
This functional equation reads:
$$F(k_1,k_2)(F(k_4,k_5)+G(k_5,k_3)+G(k_4,k_3))+F(k_1,k_3)(F(k_4,k_5)+G(k_2,k_5)+G(k_2,k_4))+$$
\begin{equation}\label{funeq}
G(k_5,k_1)(F(k_2,k_3)+G(k_4,k_3)+G(k_2,k_4))+G(k_4,k_1)(F(k_2,k_3)+G(k_5,k_3)+G(k_2,k_5))+
\end{equation}
$$F(k_4,k_5)F(k_2,k_3)+G(k_2,k_4)G(k_5,k_3)+G(k_2,k_5)G(k_4,k_3)=0$$
where $F(x,y)=\frac{x+y}{f(x+y)-f(x)-f(y)},~G(x,y)=\frac{x-y}{f(x-y)-f(x)+f(y)}$ and $k_1,...,k_5$ satisfy the following constraints:

\begin{equation}\label{con1}
k_1+k_2+k_3=k_4+k_5,
\end{equation}
\begin{equation}\label{con2}
f(k_1)+f(k_2)+f(k_3)=f(k_4)+f(k_5).
\end{equation}

Note that if $f(k)$ is a solution of this functional equation, then $f_1(k)=af(bk)+ck$ is also a solution for arbitrary constants $a,b\ne0,~c$. We call such solutions equivalent.

{\bf Proposition.} Any solution of the functional equation (\ref{funeq}) analytic near zero and such that $f(0)=0$ is equivalent to one of the following: $f_1(k)=k^2,~f_2(k)=k^3,~f_3(k)=
k^2\frac{e^k+1}{e^k-1}$. 

{\bf Remark.} Here we suppose that $a,b\ne0$ in our equivalence relation are complex numbers. If we restrict ourself to real numbers, then there exists one more non-equivalent solution 
$f_4(k)=k^2\cot(k)$.

{\bf Proof.} Set $k_2=k_4+u,~k_5=k_3+v$, then the constraint (\ref{con1}) is equivalent to $k_1=v-u$. Expanding the constraint (\ref{con2}) near $u=v=0$ we obtain 
\begin{equation}\label{v}
v=\frac{f^{\prime}(k_4)}{f^{\prime}(k_3)}u+o(u).
\end{equation} 
Expanding (\ref{funeq}) near $u=v=0$ and substituting (\ref{v}) we obtain in the first non-trivial term:
$$-(-f(k_4)+f(k_4-k_3)+f(k_3)) (f(k_4)-f(k_4+k_3)+f(k_3)) (-4 k_3 f(k_3)+2 k_3 f(k_4+k_3)-k_4 f(k_4-k_3)-$$$$k_4 f(k_4+k_3)-2 k_3 f(k_4-k_3)+2 k_4 f(k_4)) f^{\prime}(k_4)^2+$$
$$(-f(k_4)+f(k_4-k_3)+f(k_3)) (f(k_4)-f(k_4+k_3)+f(k_3))$$$$ (-2 k_3 f(k_3)+k_3 f(k_4+k_3)-2 k_4 f(k_4-k_3)-2 k_4 f(k_4+k_3)-k_3 f(k_4-k_3)+4 k_4 f(k_4)) f^{\prime}(k_3)^2+$$
$$k_4 (-f(k_4)+f(k_4-k_3)+f(k_3)) (f(k_4)-f(k_4+k_3)+f(k_3)) (2 k_4 f(k_4)+k_3 f(k_4+k_3)-k_3 f(k_4-k_3)-$$$$2 k_3 f(k_3)-k_4 f(k_4-k_3)-k_4 f(k_4+k_3)) f^{\prime}(k_3) f^{\prime\prime}(k_4)-$$
$$k_3 (-f(k_4)+f(k_4-k_3)+f(k_3)) (f(k_4)-f(k_4+k_3)+f(k_3))$$$$ (2 k_4 f(k_4)+k_3 f(k_4+k_3)-k_3 f(k_4-k_3)-2 k_3 f(k_3)-k_4 f(k_4-k_3)-k_4 f(k_4+k_3)) f^{\prime\prime}(k_3) f^{\prime}(k_4)$$
$$+(-2 k_3^2 f(k_4)^2-k_3^2 f(k_4-k_3)^2-k_3^2 f(k_4+k_3)^2+2 k_4^2 f(k_4) f(k_4+k_3)-$$$$4 k_3 k_4 f(k_4-k_3) f(k_3)+4 k_3 k_4 f(k_4) f(k_4-k_3)-2 k_4 k_3 f(k_4-k_3)^2-k_4^2 f(k_4+k_3)^2-$$
$$4 k_3 k_4 f(k_4) f(k_4+k_3)+2 k_4^2 f(k_3) f(k_4+k_3)+2 k_3^2 f(k_4+k_3) f(k_3)+8 k_4 k_3 f(k_4) f(k_3)+$$
$$2 k_3^2 f(k_4) f(k_4+k_3)+2 k_3 k_4 f(k_4+k_3)^2+2 k_3^2 f(k_4-k_3) f(k_4)-4 k_3 k_4 f(k_4+k_3) f(k_3)-k_4^2 f(k_4-k_3)^2-$$
$$2 k_4^2 f(k_4)^2-2 k_3^2 f(k_4-k_3) f(k_3)-2 k_4^2 f(k_3) f(k_4-k_3)-$$$$2 k_3^2 f(k_3)^2-2 k_4^2 f(k_3)^2+2 k_4^2 f(k_4) f(k_4-k_3)) f^{\prime}(k_4)^2 f^{\prime}(k_3)-$$
\begin{equation}
\label{funeq1}(-f(k_4)+f(k_4-k_3)+f(k_3)) (f(k_4)-f(k_4+k_3)+f(k_3)) 
\end{equation}
$$(2 k_3 f(k_3)-k_3 f(k_4+k_3)-k_4 f(k_4-k_3)-k_4 f(k_4+k_3)+k_3 f(k_4-k_3)+2 k_4 f(k_4)) f^{\prime}(k_3) f^{\prime}(k_4)+$$
$$(k_4^2 f(k_4+k_3)^2+2 k_4 k_3 f(k_4-k_3)^2+4 k_3 k_4 f(k_4-k_3) f(k_3)-2 k_4^2 f(k_4) f(k_4-k_3)+$$$$k_4^2 f(k_4-k_3)^2+k_3^2 f(k_4+k_3)^2+
2 k_3^2 f(k_4)^2+2 k_4^2 f(k_3) f(k_4-k_3)+2 k_4^2 f(k_4)^2+k_3^2 f(k_4-k_3)^2+$$$$4 k_3 k_4 f(k_4+k_3) f(k_3)-2 k_3 k_4 f(k_4+k_3)^2-2 k_3^2 f(k_4-k_3) f(k_4)+
2 k_4^2 f(k_3)^2-2 k_4^2 f(k_3) f(k_4+k_3)+$$$$4 k_3 k_4 f(k_4) f(k_4+k_3)-8 k_4 k_3 f(k_4) f(k_3)+2 k_3^2 f(k_3)^2+2 k_3^2 f(k_4-k_3) f(k_3)-$$
$$2 k_4^2 f(k_4) f(k_4+k_3)-4 k_3 k_4 f(k_4) f(k_4-k_3)-2 k_3^2 f(k_4) f(k_4+k_3)-2 k_3^2 f(k_4+k_3) f(k_3)) f^{\prime}(k_3)^2 f^{\prime}(k_4)-$$
$$k_4 (-f(k_4)+f(k_4-k_3)+f(k_3))^2 (f(k_4)-f(k_4+k_3)+f(k_3))^2 f^{\prime\prime}(k_4)+$$
$$k_3 (-f(k_4)+f(k_4-k_3)+f(k_3))^2 (k_4+k_3) f^{\prime}(k_4+k_3) f^{\prime}(k_4)^2-$$$$k_4 (-f(k_4)+f(k_4-k_3)+f(k_3))^2 (k_4+k_3) f^{\prime}(k_4+k_3) f^{\prime}(k_3)^2+$$
$$(-f(k_4)+f(k_4-k_3)+f(k_3))^2 (k_4-k_3) (k_4+k_3) f^{\prime}(k_4+k_3) f^{\prime}(k_3) f^{\prime}(k_4)-$$
$$k_3 (f(k_4)-f(k_4+k_3)+f(k_3))^2 (k_4-k_3) f^{\prime}(k_4-k_3) f^{\prime}(k_4)^2-$$
$$k_4 (f(k_4)-f(k_4+k_3)+f(k_3))^2 (k_4-k_3) f^{\prime}(k_4-k_3) f^{\prime}(k_3)^2+$$$$k_3 (-f(k_4)+f(k_4-k_3)+f(k_3))^2 (f(k_4)-f(k_4+k_3)+f(k_3))^2 f^{\prime\prime}(k_3)+$$
$$2 (-f(k_4)+f(k_4-k_3)+f(k_3))^2 (f(k_4)-f(k_4+k_3)+f(k_3))^2 f^{\prime}(k_4)-$$$$2 (-f(k_4)+f(k_4-k_3)+f(k_3))^2 (f(k_4)-f(k_4+k_3)+f(k_3))^2 f^{\prime}(k_3)+$$
$$(f(k_4)-f(k_4+k_3)+f(k_3))^2 (k_4-k_3) (k_4+k_3) f^{\prime}(k_4-k_3) f^{\prime}(k_3) f^{\prime}(k_4)=0$$

Assume without loss of generality that $f(k)=a_2k^2+a_3k^3+...$
Expanding (\ref{funeq1}) near $k_3=k_4=0$ we get $a_2a_3=0$. We have different cases:

{\bf Case 1.} Let $a_2\ne0$, then $a_3=0$. Without loss of generality we assume that $f(k)=k^2+a_4k^4+a_5k^5...$ Expanding (\ref{funeq1}) near $k_3=0$ we obtain under this assumption in 
the first non-trivial term:
$$-5 f^{\prime}(k_4)^2+8 k_4 f^{\prime\prime}(k_4) f^{\prime}(k_4)+f^{\prime}(k_4) k_4^2 f^{\prime\prime\prime}(k_4)-3 k_4^2 f^{\prime\prime}(k_4)^2=0.$$
The only solution of this 3rd-order ODE of the form $f(k)=k^2+a_4k^4+a_5k^5...$ is $f(k)=k^2$.

{\bf Case 2.} Let $a_3\ne0$, then $a_2=0$. Without loss of generality we assume that $f(k)=k^3+a_4k^4+a_5k^5...$ Expanding (\ref{funeq1}) near $k_3=0$ we obtain under this assumption in the first 
non-trivial term:
$$-4 f^{\prime}(k_4)^2 f^{\prime\prime\prime}(k_4)+12 f^{\prime}(k_4)^2-k_4 f^{\prime}(k_4)^2 f^{\prime\prime\prime\prime}(k_4)-2 f^{\prime}(k_4) 
k_4^2 f^{\prime\prime\prime}(k_4)+$$$$4 f^{\prime}(k_4) k_4 f^{\prime\prime}(k_4) f^{\prime\prime\prime}(k_4)-18 k_4 f^{\prime\prime}(k_4) f^{\prime}(k_4)+6 f^{\prime}(k_4) 
f^{\prime\prime}(k_4)^2-3 f^{\prime\prime}(k_4)^3 k_4+6 k_4^2 f^{\prime\prime}(k_4)^2=0.$$
Any solution of this 4th-order ODE of the form $f(k)=k^3+a_4k^4+a_5k^5...$ is equivalent to either $f(k)=k^3$ or $f(k)=k^2\frac{e^k+1}{e^k-1}$  (or $f(k)=k^2\cot(k)$ if our 
group of equivalence is real rather then complex).

Note that in the case $a_2=a_3=0$ there are not non-trivial solutions.

\section{New equations}
\bigskip

Our results show that if we set
$$F(k)=ak^3\cot bk-ck$$
where $a,~b,~c$ are constants, then the five-wave amplitude is also zero. The corresponding equation hardly has any physical importance because $F(k)=\infty$ at $bk=\pi n$. In any case, in the limit 
$b\to 0,~a=\frac{3}{b},~c=\frac{3}{b^2}$ we get $F(k)=k^3$ and the equation goes to the KdV. Another equation is more interesting. Let $u=u(x,y,t)$ be a function in 
two spatial coordinates $x,~y$ and $F$ is given by
$$F(\frac{\partial}{\partial x},\frac{\partial}{\partial y})=a\frac{\partial^2}{\partial x^2}\coth a\frac{\partial}{\partial y}.$$
The five-wave amplitude is zero in this case. This equation may be useful in applications. In the limit $a\to 0$ we have 
$F\to \frac{\partial^2}{\partial x^2}\Big(\frac{\partial}{\partial x}\Big)^{-1}$ and our equation degenerates to the following form
$$u_t=\partial_x^2\partial_y^{-1}u+uu_x.$$
This equation can be compared with Khokhov-Zabolotskaya equation (see, for instance, \cite{z6})
$$u_t=\partial_y^2\partial_x^{-1}u+uu_x.$$
But these equations are not equivalent.

\vskip.3cm \noindent {\bf Aknowledgements} 
This work was supported by the NSF grant NSF40CE-1130450 and the Russian Government contract 11.934.31.0035 (signed November 25, 2010).

\end{document}